# A Novel Taxonomy for Navigating and Classifying Synthetic Data in Healthcare Applications

Bram VAN DIJK [a], Saif ul ISLAM [b], Jim ACHTERBERG [a], Hafiz MUHAMMAD WASEEM [b], Parisis GALLOS [c,1], Gregory EPIPHANIOU [b], Carsten MAPLE [b], Marcel HAAS [a] and Marco SPRUIT [a]

[a] *Leiden University Medical Center, The Netherlands*
[b] *University of Warwick, UK*
[c] *European Federation for Medical Informatics, Switzerland*

ORCiD ID: Bram van Dijk https://orcid.org/0009-0002-9176-1608,
Saif ul Islam https://orcid.org/0000-0002- 9546-4195,
Jim Achterberg https://orcid.org/0009-0000- 9589-7831,
Hafiz Muhammad Waseem https://orcid.org/0000-0002-9418-1492,
Parisis Gallos https://orcid.org/0000-0002-8630-7200,
Gregory Epiphaniou https://orcid.org/0000-0003-1054-6368,
Carsten Maple https://orcid.org/0000-0002-4715-212X,
Marcel Haas https://orcid.org/0000-0003- 2581-8370,
Marco Spruit https://orcid.org/0000-0002-9237-221X

**Abstract.** Data-driven technologies have improved the efficiency, reliability and effectiveness of healthcare services, but come with an increasing demand for data, which is challenging due to privacy-related constraints on sharing data in healthcare contexts. Synthetic data has recently gained popularity as potential solution, but in the flurry of current research it can be hard to oversee its potential. This paper proposes a novel taxonomy of synthetic data in healthcare to navigate the landscape in terms of three main varieties. Data Proportion comprises different ratios of synthetic data in a dataset and associated pros and cons. Data Modality refers to the different data formats amenable to synthesis and format-specific challenges. Data Transformation concerns improving specific aspects of a dataset like its utility or privacy with synthetic data. Our taxonomy aims to help researchers in the healthcare domain interested in synthetic data to grasp what types of datasets, data modalities, and transformations are possible with synthetic data, and where the challenges and overlaps between the varieties lie.

## 1. Introduction

Digital technologies have transformed economies and societies and hold a lot of potential for the healthcare domain. Health technologies like the Electronic Health Record (EHR) and Clinical Decision Support Systems [CDSS] are estimated to yield high benefits in terms of gains in cost efficiency, reliability, and effectiveness of health services [1].

---

[1] Corresponding Author: Parisis Gallos; E-mail: parisgallos@yahoo.com.

These technologies require data in high volume and quality, but due to the presence of sensitive personal information in medical data, strict privacy regulations such as the General Data Protection Regulation (GDPR) are applicable to its sharing. Consequently, limited medical data is available for analysis and research, which hampers innovation [1, 2]. Hence, researchers and practitioners are turning towards Synthetic Data (SD), which can be defined as '*data that has been generated using a purpose-built mathematical model or algorithm, with the aim of solving a (set of) data science task(s)*' [2]. Properly generated SD ensures enhanced *privacy* as it does not refer to real individuals, maintains good *fidelity* by replicating statistical patterns found in Real Data (RD), and offers *utility* by effectively substituting real data in task-specific applications.

Interest in synthetic data is surging [3], but comprehending the full potential of SD in healthcare contexts remains challenging. This paper proposes a novel taxonomy of SD within digital healthcare settings that allows researchers and medical professionals interested in exploring SD to grasp:

- what types of datasets SD renders possible;
- what data modalities are amenable to SD;
- what kinds of dataset transformations are enabled by SD.

In the remainder of this paper, we describe our taxonomy and varieties in Section 2, and reflect on our findings in Section 3.

## 2. Synthetic Data Taxonomy

Figure 1 presents the proposed taxonomical classification of SD. We categorize SD into three main varieties: Data Proportion, Data Modality, and Data Transformation.

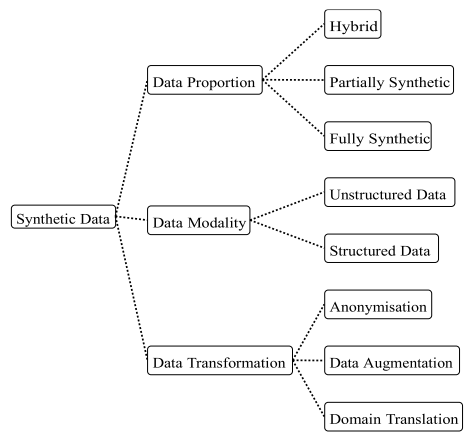

**Figure 1.** Taxonomy of SD varieties relevant to healthcare applications.

*2.1. Data Proportion*

Data Proportion refers to the ratio of SD to total data in a dataset, where SD can comprise additional variables or observations in the dataset. Data Proportion indirectly bears on privacy risks associated with the SD. We classify Data Proportion further into *fully SD*, *partially SD*, and *hybrid SD*, following [1].

Fully SD is entirely generated by an algorithm or mathematical model and contains no RD. However, information from RD can potentially leak into the synthetic dataset [1]. Examples of fully SD in healthcare include synthetic EHRs [4], medical images [5,6], and clinical notes [7].

Partially SD consists of a combination of SD and RD within a single observation. Here SD usually replaces sensitive information, while non-sensitive information remains unchanged [2]. Examples of partially SD in healthcare include partially synthetic EHRs [8], health surveys [9], and clinical notes [10].

Hybrid SD contains both entirely synthetic and entirely real observations. Since entirely new observations have to be synthesized, here the same methods are employed as in fully SD. Examples of hybrid SD in healthcare are dataset augmentation in EHRs [11], medical images [12], and clinical notes [13].

Each Data Proportion type has unique benefits and limitations. Fully SD can address data scarcity issues and ethical, legal, or privacy-related constraints. However, since fully SD mimics statistical patterns found in RD, privacy risks may still exist [14]. Partially SD is useful when RD is abundant, but sensitive information needs to be protected. However, since observations in partially SD may still be directly traceable to real individuals, privacy risk may be higher as well. Hybrid SD can obfuscate the identifiability of individual records while maintaining data utility, but as hybrid SD also contains real observations, privacy risks are often larger than for fully and partially SD.

*2.2. Data Modality*

We categorize Data Modality based on whether data are stored in a *structured* or *unstructured* format. Structured data contains predefined variables or features of interest [15], whereas in unstructured data additional processing is required to extract features.

Structured data is organized in a specific format such as tables, where rows represent observations or samples and columns denote variables or features. In the medical domain, various types of structured data are common: *tabular data* (e.g., EHRs), *time series data* (e.g., blood pressure and heart rate over time), and *graph data* (e.g., a network of diseases). Generating structured data presents specific challenges, particularly concerning privacy. For example, in the case of EHRs, strong correlations between SD and RD observations can potentially allow an attacker to infer sensitive features or identity information about an individual. In addition, tabular data like EHRs often include heterogeneous features with vastly different properties and distributions, making it challenging to represent them collectively [17].

Unstructured data consists of various formats such as *text* in clinical notes, radiological *images*, ultrasound *videos*, phonocardiographic *audio*, and combinations of these in multi-modal formats (for example pairs of radiological images and text labels). Current challenges in generating unstructured SD involve adequately assessing its fidelity. For example, popular fidelity metrics for synthetic images may poorly capture desired statistical properties of medical images, resulting in biased evaluation [16]. Also, models generating medical images often rely on architectures and loss functions designed for generating natural images, which may induce spurious artefacts in the SD [16].

*2.3. Data Transformation*

SD can also be used to modify existing datasets, so as to improve their representation of a population (fidelity), downstream task performance (utility), and protection against inference (privacy). We classify health data transformation into three main categories: *anonymization*, *data augmentation*, and *domain translation*.

Anonymization means replacing sensitive features in a dataset with SD, to reduce the risk that other sensitive features or a real individual itself can be inferred from the SD. Using SD for anonymization has been done in EHRs [8], health surveys [9], and clinical notes [10].

Data Augmentation refers to using SD to improve dataset utility and/or fidelity. Examples are generating synthetic samples of an underrepresented group to improve the performance of a classifier for that group, or adding these samples to a dataset to improve its representativeness overall. Data augmentation has been employed to improve EHRs [11], medical images [12], and clinical notes [13] regarding utility and/or privacy.

Domain Translation refers to the transformation of data within a particular modality and domain. Examples are the synthesis of Magnetic Resonance images from Computer Tomography images, where the former are costly to obtain [6], and the synthesis of high- from low-resolution Magnetic Resonance images [5].

## 3. Discussion and Conclusion

This paper provided a taxonomy of SD in healthcare applications. We categorized the SD based on the proportion of SD in a dataset, data modality, and types of dataset transformations that SD supports. As an illustration of its practical value, we envision integration of the taxonomy as a roadmap in more general frameworks and toolkits. An example is the INSAFEDARE project, where it can guide researchers, developers and regulatory bodies in employing SD to more efficiently obtain quality and safety assurance of various digital health applications [18].

It is important to note that the three varieties of SD are interrelated and applicable across various applications of SD. Hybrid SD, for example, performs data augmentation as data transformation and can consist of structured or unstructured data. Partially SD often performs anonymization as data transformation and is usually structured rather than unstructured data. Sensitive features are more easily identified in structured than unstructured data. Lastly, fully SD can be structured or unstructured and can be used for domain translation, augmentation, or anonymization.

## Acknowledgments

This work is co-funded by the HORIZON.2.1 - Health Programme of the European Commission, Grant Agreement number: 101095661 - Innovative applications of assessment and assurance of data and synthetic data for regulatory decision support (INSAFEDARE).

## References

[1]   Murtaza H, Ahmed M, Khan NF, Murtaza G, Zafar S, Bano A. Synthetic data generation: State of the art in health care domain. Computer Science Review. 2023;48(100546).
[2]   Jordon J, Szpruch L, Houssiau F, Bottarelli M, Cherubin G, Maple C, et al. Synthetic Data – what, why and how? arXiv preprint arXiv:220503257. 2022.
[3]   Jordon J, Wilson A, van der Schaar M. Synthetic data: Opening the data floodgates to enable faster, more directed development of machine learning methods. arXiv preprint arXiv:201204580. 2020.
[4]   Yan C, Yan Y, Wan Z, Zhang Z, Omberg L, Guinney J, et al. A multifaceted benchmarking of synthetic electronic health record generation models. Nature communications. 2022;13(1):7609.
[5]   Nie D, Trullo R, Lian J, Wang L, Petitjean C, Ruan S, et al. Medical image synthesis with deep convolutional adversarial networks. IEEE Transactions on Biomedical Engineering. 2018;65(12):2720-30.
[6]   Yang H, Lu X, Wang SH, Lu Z, Yao J, Jiang Y, et al. Synthesizing Multi-Contrast MR Images Via Novel 3D Conditional Variational Auto-Encoding GAN. Mobile Networks and Applications. 2021;26:415-24.
[7]   Li J, Zhou Y, Jiang X, Natarajan K, Pakhomov SV, Liu H, et al. Are synthetic clinical notes useful for real natural language processing tasks: A case study on clinical entity recognition. Journal of the American Medical Informatics Association. 2021;28(10):2193-201.
[8]   Zhou N, Wang L, Marino S, Zhao Y, Dinov ID. DataSifter II: Partially synthetic data sharing of sensitive information containing time-varying correlated observations. Journal of Algorithms & Computational Technology. 2022;16:17483026211065379.
[9]   Loong B, Zaslavsky AM, He Y, Harrington DP. Disclosure Control Using Partially Synthetic Data for Large-Scale Health Surveys, with Applications to CanCORS. Statistics in medicine. 2013;32(24):4139-61.
[10]  Zhou N, Wu Q, Wu Z, Marino S, Dinov ID. DataSifterText: Partially Synthetic Text Generation for Sensitive Clinical Notes. Journal of Medical Systems. 2022;46(12):96.
[11]  Che Z, Cheng Y, Zhai S, Sun Z, Liu Y. Boosting deep learning risk prediction with generative adversarial networks for electronic health records. In: 2017 IEEE International Conference on Data Mining (ICDM). IEEE; 2017;787-92.
[12]  Chen Y, Yang XH, Wei Z, Heidari AA, Zheng N, Li Z, et al. Generative adversarial networks in medical image augmentation: a review. Computers in Biology and Medicine. 2022;144:105382.
[13]  Latif A, Kim J. Evaluation and Analysis of Large Language Models for Clinical Text Augmentation and Generation. IEEE Access. 2024;12:48987-96.
[14]  Abowd JM, Vilhuber L. How protective are synthetic data? In: International Conference on Privacy in Statistical Databases. Springer; 2008. p. 239-46.
[15]  Boulton D, Hammersley M. Analysis of Unstructured Data. Data collection and analysis. 2006;2:243-59.
[16]  Ibrahim M, Khalil YA, Amirrajab S, Suna C, Breeuwer M, Pluim J, et al. Generative AI for Synthetic Data Across Multiple Medical Modalities: A Systematic Review of Recent Developments and Challenges. arXiv preprint arXiv:240700116. 2024.
[17]  Yoon J, Mizrahi M, Ghalaty NF, Jarvinen T, Ravi AS, Brune P, et al. EHR-Safe: generating high-fidelity and privacy-preserving synthetic electronic health records. NPJ Digital Medicine. 2023;6(1):141.
[18]  Gallos P, Matragkas N, Islam SU, Epiphaniou G, Hansen S, Harrison S, et al. INSAFEDARE Project: Innovative Applications of Assessment and Assurance of Data and Synthetic Data for Regulatory Decision Support. Studies in Health Technology and Informatics. 2024;316:1193-97.